\documentclass[letter,twocolumn]{jpsj2} %% two-column layout

\title{Clustering Analysis of Periodic Point Vortices with the $L$ Function}

\author{\textsc{Makoto Umeki}\thanks{E-mail address: umeki@phys.s.u-tokyo.ac.jp}}

\inst{Department of Physics, 
Graduate School of Science, 
University of Tokyo, \\
7-3-1 Hongo, 
Bunkyo-ku, Tokyo 113-0033, Japan}

\abst{
The motion of point vortices with periodic 
boundary conditions was studied by using 
Weierstrass zeta functions. 
The scattering and recoupling of a vortex pair by a 
third vortex becomes remarkable when the vortex 
density is large. 
The clustering of vortices with various initial conditions 
is quantitated by the $L$ function used in 
the point process theory in spatial ecology. 
It is shown that clustering persists if the initial distribution is 
clustered like an infinite row or a checkered pattern. }

\kword{point vortex, two-dimensional turbulence, 
$L$ function, point process theory}

\begin{document}
\maketitle

The statistical approach to the problem of assemblies of point 
vortices (PVs) dates back to Onsager (1949). A state of negative 
temperature is considered to be related to the clustering of 
vortices rotating in the same direction and 
the inverse energy cascade predicted in the two-dimensional 
Navier-Stokes (2D NS) turbulence. 
In many numerical simulations, PVs are 
bounded in a circular wall since a velocity field 
due to a PV can be computed by 
including a single mirror image. 
Although the axisymmetry with respect to the origin 
is conserved, spatial homogeneity is not guaranteed 
in such a circular system. 
The numerical difficulty in the simulation 
of vortices in a box is that there emerges an 
infinite sequence of virtual images. 
Our objective is to study the turbulent motions and 
clustering of many PVs in a periodic box using 
Weierstrass zeta functions. 

Let us start by representing the 2D NS equation 
in terms of a complex 
position $z=x+iy$, velocity $q=u-iv$, pressure $p$, 
and kinematic viscosity $\nu$ as 
\begin{equation}
q_t + q q_{\bar{z}}+ \bar{q} q_z = 
-2 p_z + 4 \nu q_{z\bar{z}}.
\label{eq1}
\end{equation}
Here, $\bar{q}$ denotes the complex conjugate of $q$ and  
we use the relations $\partial_x=\partial_z+\partial_{\bar{z}},$ 
$\partial_y=i(\partial_z-\partial_{\bar{z}}),$ $u=(q+\bar{q})/2,$ 
$v=i(q-\bar{q})/2$, and 
$\Delta=4 \partial_{z\bar{z}}$.
The incompressible condition yields
$\nabla \cdot {\mathbf v} = \bar{q}_{z}+q_{\bar{z}}=0$.
The vorticity $\omega= v_x-u_y$ can be expressed by $q$ 
as $ \omega= 2iq_{\bar{z}}$.

If the flow is irrotational $\omega =0$, then $q_{\bar{z}}=0$, 
$q$ depends on only $z$ (and $t$), and the theory of conformal 
mapping can be applied. 
Equations for the vorticity and pressure are
\begin{equation}
q_{\bar{z}t} + q q_{\bar{z}\bar{z}} + \bar{q} q_{z\bar{z}}
= 4\nu q_{z\bar{z}\bar{z}},
\label{eq4}
\end{equation}
and $p_{z\bar{z}} = -(q_z \bar{q}_{\bar{z}}+ \bar{q}_z^2)/2$, 
respectively.

According to Tkachenko\cite{tka1,tka2}, 
the velocity field due to a single PV at the origin with periodic 
boundary conditions (BCs) is equivalent to that due to the PVs 
on the lattice $z_{mn}= 2m\omega_1\ +\ 2n\omega_2$, 
where the complex numbers $\omega_1, \omega_2$ are 
the half periods of the lattice and $m,n$ are arbitrary integers.  
The ratio of the two periods $\tau= \omega_1/\omega_2$ 
can be restricted in the region
\begin{equation}
{\rm Im} \tau >0 ,\ \  |{\rm Re} \tau| <1/2, \ \  |\tau| \ge 1.
\label{eq6}
\end{equation}

We investigate the case of square periodic BCs, 
which are usually employed in the numerical studies of 
two-dimensional turbulence by selecting $\tau=i$. 
However, we can deal with an arbitrary periodic 
parallelogram by considering various values of $\tau$ 
that satisfy (\ref{eq6}). 

The velocity field due to a PV of strength $\kappa=2\pi$ 
is given by the Weierstrass zeta function 
$\zeta(z;\omega_1,\omega_2)$ 
along with a rigid rotation term as follows:
\begin{equation}
\bar{q} = i \overline{\zeta(z)} - i \Omega z \equiv w(z). 
\label{eq7}
\end{equation}
Since the vortex lattice undergoes rigid rotation with 
an angular velocity 
$\Omega = \pi/[ 4 {\rm Im} (\bar{\omega}_1 \omega_2) ]$,  
the second term in Eq. (\ref{eq7}) is necessary in order to cancel 
the velocity circulation on the boundary. 
The vortex density $n=1/[4 {\rm Im}  (\bar{\omega}_1 \omega_2) ]$, 
$\Omega$, and the vortex strength $\kappa$ 
are related as $\kappa n = 2\Omega$. 
If the length of the side of the square is unity, then 
$\omega_1=1/2$, $\omega_2=i/2$, $\Omega=\pi$, and $\kappa=2\pi$. 

The equation for the streamline $\psi={\rm const.}$, 
where $\psi$ is the streamfunction, 
is equivalent to $dx/\psi_y= -dy/\psi_x$. 
Using the relations $u=\psi_y$ and $v=-\psi_x$, \ 
$\psi$ is expressed as \ 
$\psi=\int u dy \ +\ f(x)$.
The sigma and zeta functions of Weierstrass are related as
$\zeta(z)=\sigma'(z)/\sigma(z)$. 
This relation is consistent with the asymptotic forms 
$\zeta\sim 1/z$ and $\sigma \sim z$ when $z\sim 0$. 
Using the above results, $\psi$ for a single vortex lattice 
centered at the origin is given by
\begin{equation}
\psi=-{\rm Re} \ln \sigma(z) +\Omega |z|^2 /2.
\label{eq12}
\end{equation}
There is a minimum value of $\psi$ in the periodic case, 
in contrast with the unbounded plane in which there are no limits on $\psi$ .

For simplicity, we consider an assembly of PVs with $\kappa_i= 2\pi \mu_i$, 
$\mu_i=1,$ for $i=1, \cdots, N_1$ and 
$\mu_i=-1$ for $i=N_1 +1, \cdots, \ N(=N_1 + N_2)$.  
Therefore, $\psi$ for $N$ PVs located at $z_i$ is given by 
\begin{equation}
\psi=\Sigma_{i=1}^N \mu_i 
\{ -{\rm Re}[ \ln \sigma(z-z_i)] + \Omega |z-z_i|^2 /2 \}. 
\label{eq13}
\end{equation}
Using Eq. (\ref{eq7}), the equation of motion of PVs  
with square periodic BCs can be expressed as\cite{Aref1999}
\begin{equation}
\dot{z}_i= \Sigma_{j\ne i} \mu_j w(z_i-z_j).   
\label{eq14}
\end{equation}
The equation can be rewritten in the Hamiltonian form as 
$\mu_i dz_i/dt= \partial H/\partial \bar{z}_i$, 
where the Hamiltonian $H$ can be expressed as 
$ H= \sum_{i=1}^{N}\mu_i h_i $ 
$=\sum_{i=1}^N\sum_{j=i+1}^N \mu_i \mu_j  h_{ij}$,
$h_{ij}= $
$
- {\rm Re}[ \ln \sigma(z_i-z_j)]$
$\  +\  \Omega |z_i-z_j|^2/2$. 
The Hamiltonian, 
which is given by the total kinetic energy minus the self-induced 
kinetic energy of PVs, 
can be interpreted as the sum of a kinetic energy due 
to interactions between the pairs of PVs. 

If {\it Mathematica} is used, we can compute the 
Weierstrass zeta function as we use the sinusoidal 
function in a Fortran code. 
The system (\ref{eq7},\ref{eq14}) is solved numerically 
by the {\it NDSolve} command of {\it Mathematica} 5.2 
installed in a PC with an AMD Athlon 64x2 3800 CPU, 
2 GB memory and Windows XP OS. 
The computation is realistic since the CPU time in such 
a PC environment ranges from two to five days 
for 100 PVs and 10 eddy turnover times.

If the PVs lie in an unbounded domain, 
the system has four integrals:\cite{Saffman} the 
Hamiltonian $H_u=-\sum \mu_i \mu_j \ln|z_i-z_j|$, 
two components of the linear impulse ${\bf I}=(I_x,I_y)$, 
$I_x=\sum \mu_i {\rm Re}[z_i]$, 
$I_y=\sum \mu_i {\rm Im}[z_i]$,  
and the angular impulse $A= \sum \mu_i |z_i|^2$. 
Since the system in a periodic box has no circular symmetry,  
$A$ is no more constant; 
however, $H, I_x$, and $I_y$ are conserved. 
Since there are three conserved quantities, the system of 
three PVs is integrable, while the four PVs 
exhibit chaos. 

Examples of the trajectories of three PVs with $\mu_i=(2,2,-1)$ 
and an initial condition $(z_1,z_2,z_3)$ $=(0, 0.5, $ $0.25+ i \sqrt{3}/4)$ 
located at the vertices of an equilateral triangle 
and that of four PVs with $\mu_i=(2,2,-1,-1)$ 
and $(z_1,z_2,z_3,z_4)$ $=(0, 0.5,i \sqrt{3}/4,$ $ 0.5+i\sqrt{3}/4)$ 
at $t=0$ located at the vertices of a rectangle are shown 
in Figures 1a and 1b, respectively. 
The first case leads to a collapse if the PVs are 
in an unbounded plane. 

%Fig.1
%\if0
\begin{figure}
\begin{center}
\scalebox{0.26}{\includegraphics{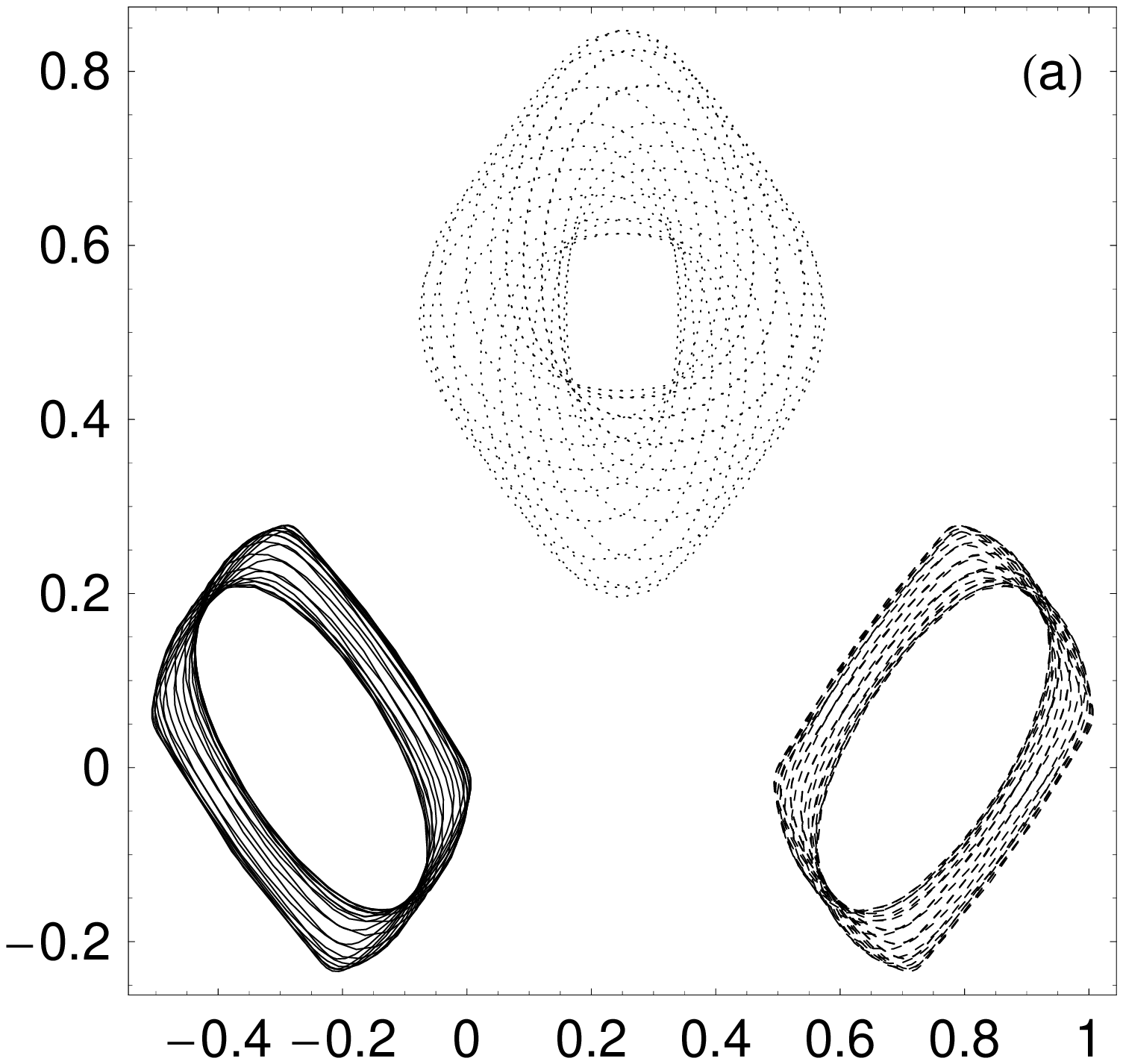}}
\scalebox{0.26}{\includegraphics{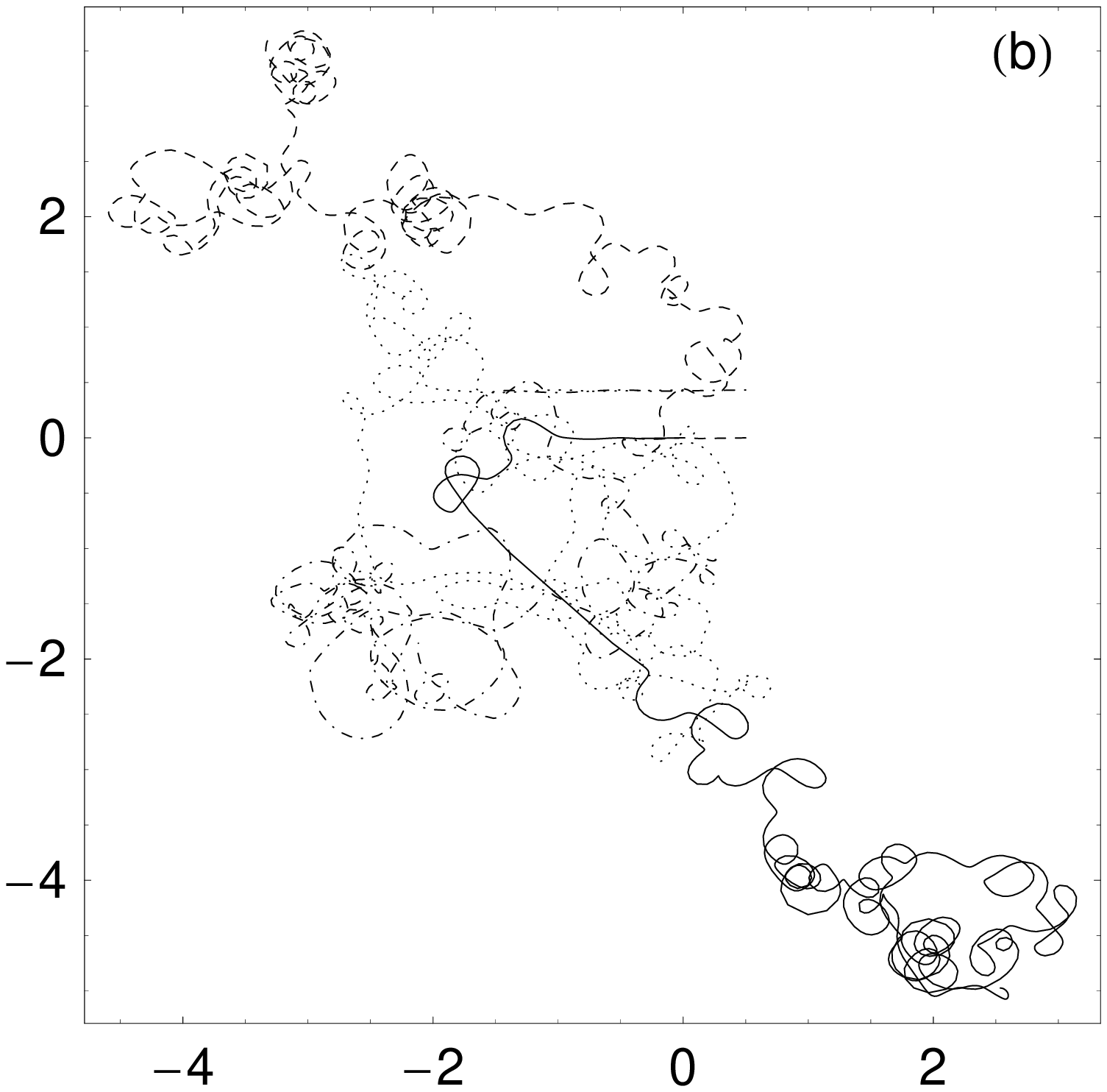}}
\end{center}
\caption{Trajectories of three (a) and four PVs (b). 
PVs 1, 2, 3, and 4 are denoted by solid, dashed, 
dotted, and dotted-dashed curves, respectively.}
\end{figure}
%\fi
%Fig.1

To analyze the spatial distribution of many PVs, 
we introduce the $L$ function used in the point process theory in 
spatial ecology\cite{Cressie,Shimatani}. 
Let ${\vec{x}_i}=(x_i,y_i)$ be the position of $N$ points randomly 
distributed in an area $S$. The $K$ function is defined by 
\begin{equation}
K(r) = (\lambda N)^{-1}
\Sigma_{i=1}^N\Sigma_{j=1,j\ne i}^N
\theta(r - | \vec{x}_i-\vec{x}_j | ), 
\label{eq01}
\end{equation}
where $\lambda=N/S$ is the number density of the points and 
$\theta(x)$ is the step function. 
An extra function added in (\ref{eq01}), in order to modify the 
edge effect, is unnecessary in the present periodic case.

If the distribution of points obeys completely spatially randomness 
(CSR), which is synonymous with a homogeneous Poisson process, 
$K(r)$ becomes the area of the circle with 
radius $r$, i.e., $K(r)=\pi r^2$. 
Then, it is convenient to introduce the $L$ function as 
\begin{equation}
L(r)= \sqrt{K(r)/\pi} -r.
\label{eq02}
\end{equation}
CSR yields $L=0$. For clustering, i.e. points staying 
close to the other points, we have $L>0$. 
If the points tend to be at a distance from each other, $L<0$. 
Whether $L(r)$ is positive or negative can depend on $r$. 
For instance, a checkered pattern ({\it Ichimatsu moyo} in Japanese) 
gives $L>0$ for small $r$, but $L<0$ for large $r$. 

According to Novikov (1976)\cite{Novikov}, for an 
unbounded plane, we have a relation between 
the distance $r_{jl}$ of two PVs 
of strength $\kappa_j$ and $\kappa_l$ 
and the energy spectrum $E(k)$ as
\begin{equation}
E(k)= (4\pi k)^{-1} 
[ \Sigma_j \kappa_j^2 + 2 \Sigma_{j<l} \kappa_j\kappa_l J_0(k r_{jl})].
\label{eq03}
\end{equation}
If $\kappa_j=\kappa$ for all $j$, we have
\begin{equation}
E(k)= \kappa^2(4\pi k)^{-1} 
[ N + 2 \Sigma_{j<l} J_0(k r_{jl})].
\label{eq04}
\end{equation}
Using the number density $\rho(r)$ at the distance $r$ 
between two PVs, we have, in the continuous limit, 
$E(k)= E_1(k)+E_2(k)$,
where $E_1(k)=\kappa^2 N (4\pi k)^{-1}$ and
$ E_2(k)=\kappa^2 (2\pi k )^{-1} \int_0^\infty J_0(k r) \rho(r) dr$. 
$E_1(k)$ and $E_2(k)$ correspond to the self-energy of 
each vortex and the interaction energy between two PVs, 
similar to $h_{ij}$. 
The relation between the $K(r)$ and $\rho(r)$ is
$
K(r)= l^2 N'^{-1} \int_0^r \rho(r') dr',$
where $\rho(r)$ is normalized by the upper limit $l$ of $r$ 
and the total number $N'=N(N-1)/2$ of pairs of PVs. 

If $\rho(r)=C r^\alpha$, we can integrate $E_2(k)$ into
$E_2(k)=$
$ \kappa^2 C k^{-\alpha-2} $
$\Gamma((\alpha+1)/2)$ 
$[2\pi \Gamma((1-\alpha)/2)]^{-1}$ for $-1<\alpha<1/2$. 
The value $\alpha=-1/3 $ gives the Kolmogorov spectrum $k^{-5/3}$ 
in a 3D turbulence. 

Formally, the CSR value $\alpha=1$ in two dimensions yields 
the $k^{-3}$ spectrum, although the integral does not converge.
In order to avoid divergence, an exponential decay 
in $\rho(r)$ may be introduced. 
Otherwise, we may set the upper limit $l$ in the range of 
integration in $E_2(k)$. 
Using the normalization $r=l r'$ and the integral expressed by 
a regularized hypergeometric function as 
$\int_0^1 J_0(kr)$ $ r^\alpha dr=$
$\Gamma((1+\alpha)/2)$
$ _1 F_2((1+\alpha)/2;$ 
$\{1,(3+\alpha)/2\};-k^2/4)$
$[ 2\Gamma((3+\alpha)/2)]^{-1}$,
for $\alpha>-1$, $E_2(k)$ is written as
$E_2(k)= $
$\kappa^2 C l^{\alpha+1}$
$ \Gamma((1+\alpha)/2)$
$ _1 F_2((1+\alpha)/2;\{1,(3+\alpha)/2\}$
$;-(kl)^2/4)$
$[ 4\pi k \Gamma((3+\alpha)/2)]^{-1}$.

If we use the CSR distribution $\rho(r)= \pi N' l^{-2} r$, 
we have $\int_0^1 J_0(kr) r dr = J_1(k)/k$.
Then, $E_2(k)$ is given by 
\begin{equation}
E_2(k)= \kappa^2  N' l^{-1} k^{-2} J_1(kl)/2.
\label{eq013}
\end{equation}
For large $kl$, we have the asymptotic form 
 \begin{equation}
E_2(k) \simeq \sqrt{1/2\pi} \kappa^2 N' l (kl)^{-5/2} \cos(kl-3\pi/4).
\label{eq014}
\end{equation}
The value of $E_2(k)$ is oscillatory as $k$ increases 
with the amplitude decaying as $k^{-5/2}$. 
Evidently, the total energy spectrum does not 
become negative because $E_1(k)\gg E_2(k)$ for large $k$. 

For numerical studies, we first 
consider an assembly of PVs having the same positive 
strength $\kappa$\ (= \ 2$\pi$). The following four typical cases are considered: 
Case (I) an infinite row thatis a discrete model of the vortex sheet, 
Case (II) PVs located randomly in checkered patterns,  
Case (III) PVs located randomly in the 10 $\times$ 10 subsquares,
and Case (IV) CSR in the unit square. 
Here, the word {\it CSR} implies that the PVs are distributed by 
using random numbers generated by a single run. 
The initial conditions, 
the number $N$ of PVs, the final time $t_f$, and the values of 
three conserved quantities are summarized in Table 1.
The relative precision of $H$ in the numerical simulation 
is confirmed to be less 
than $10^{-6} \sim 10^{-5}$ up to $t=t_f$. 
Figure 2 shows the $L$ function computed by the initial and 
final distributions of PVs. 

For Case (I), the PVs are initially located on the $x$-axis as 
$ z_{j}(0)= j/N +$ 
$ \epsilon \sin 2\pi j/N,$ 
$j=1, \cdots , N,$
where $\epsilon=0.05$. 
If $\epsilon$ is fixed and $N$ is increased, the pairing 
of two adjacent PVs becomes more conspicuous 
than the winding of the sheet due to the Kelvin-Helmholtz 
instability. The growth rate $\sigma = \kappa \pi p(1-p)/a^2$ 
of the pairing instability in an unbounded plane 
attains a maximum at the wavenumber $p=1/2$, where 
$a$ is the distance between two adjacent PVs.\cite{Saffman}
Modulus 1 is suitably considered so that the PVs are 
plotted in the selected square. 
Of course, the PVs wander chaotically from one square 
to another. 

For $\epsilon=0$, we have $\rho(r)=2N'l^{-2},$ $K(r)=2r$ 
and the average Hamiltonian 
$ \langle h \rangle =2 \sum_{i>j} h_{ij} [N(N-1)]^{-1}$
$=2[\sum_{j=1}^{N/2-1}$
$\psi(j/N)+$$\psi(1/2)/2]$$[N(N-1)]^{-1}$
$\sim 2\int_0^{1/2} dx \psi(x)$. 
Since $\psi(x)\sim -\ln |x|$ for $x\sim 0$, 
$\langle h \rangle \equiv \int_S dx dy \psi(x,y) \rho(r)$ 
yields a finite value for $\rho =C r^\alpha$, $r \sim 0$, 
and $\alpha> -2$. 
Since $\alpha <1$ corresponds to clustering, clustering 
with $\alpha<-2$ cannot occur from the initial condition 
with a finite $\langle h \rangle$. 

%\if0
%Fig.2
\begin{figure}
\begin{center}
\scalebox{0.33}{\includegraphics{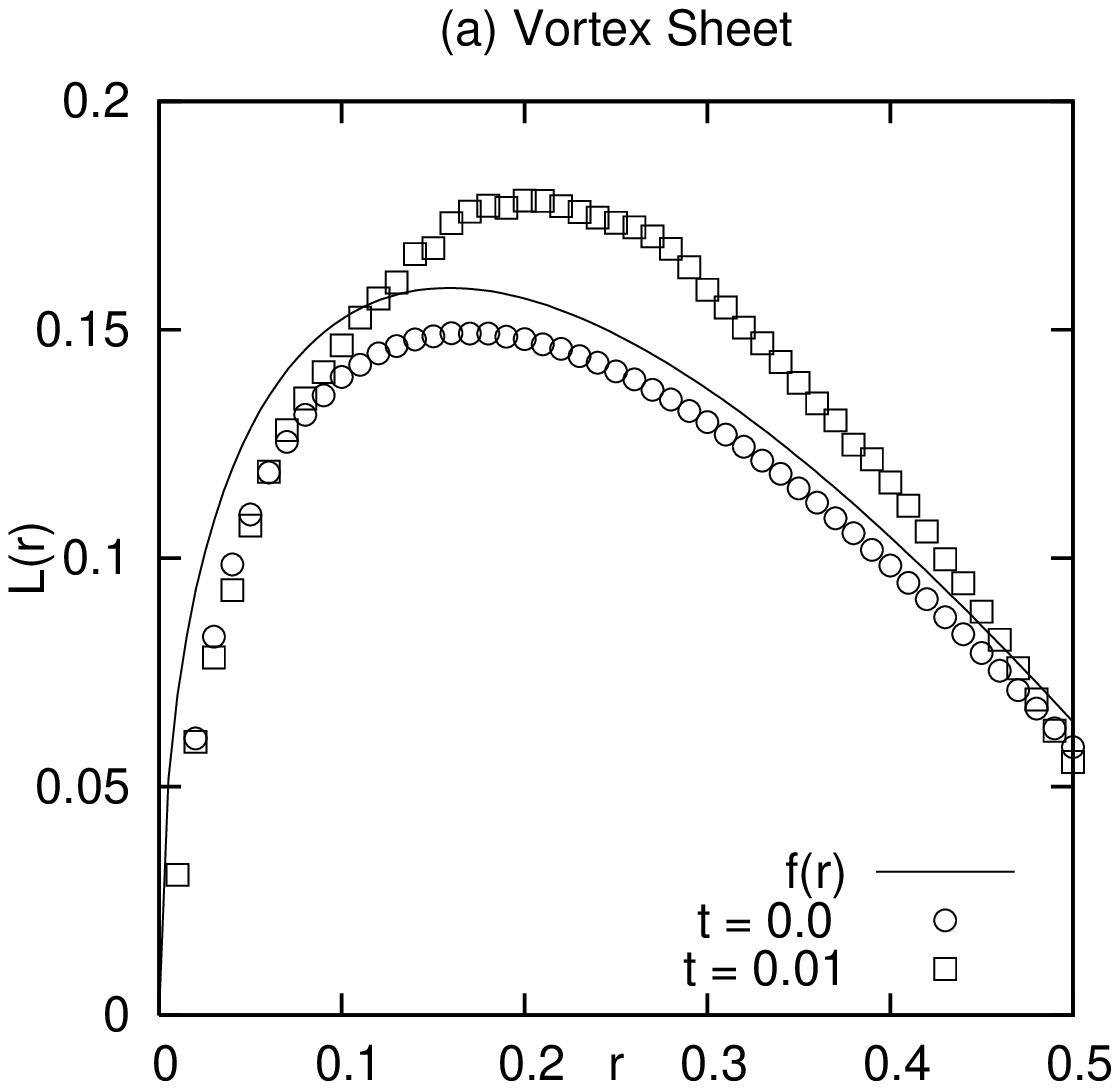}}
\scalebox{0.33}{\includegraphics{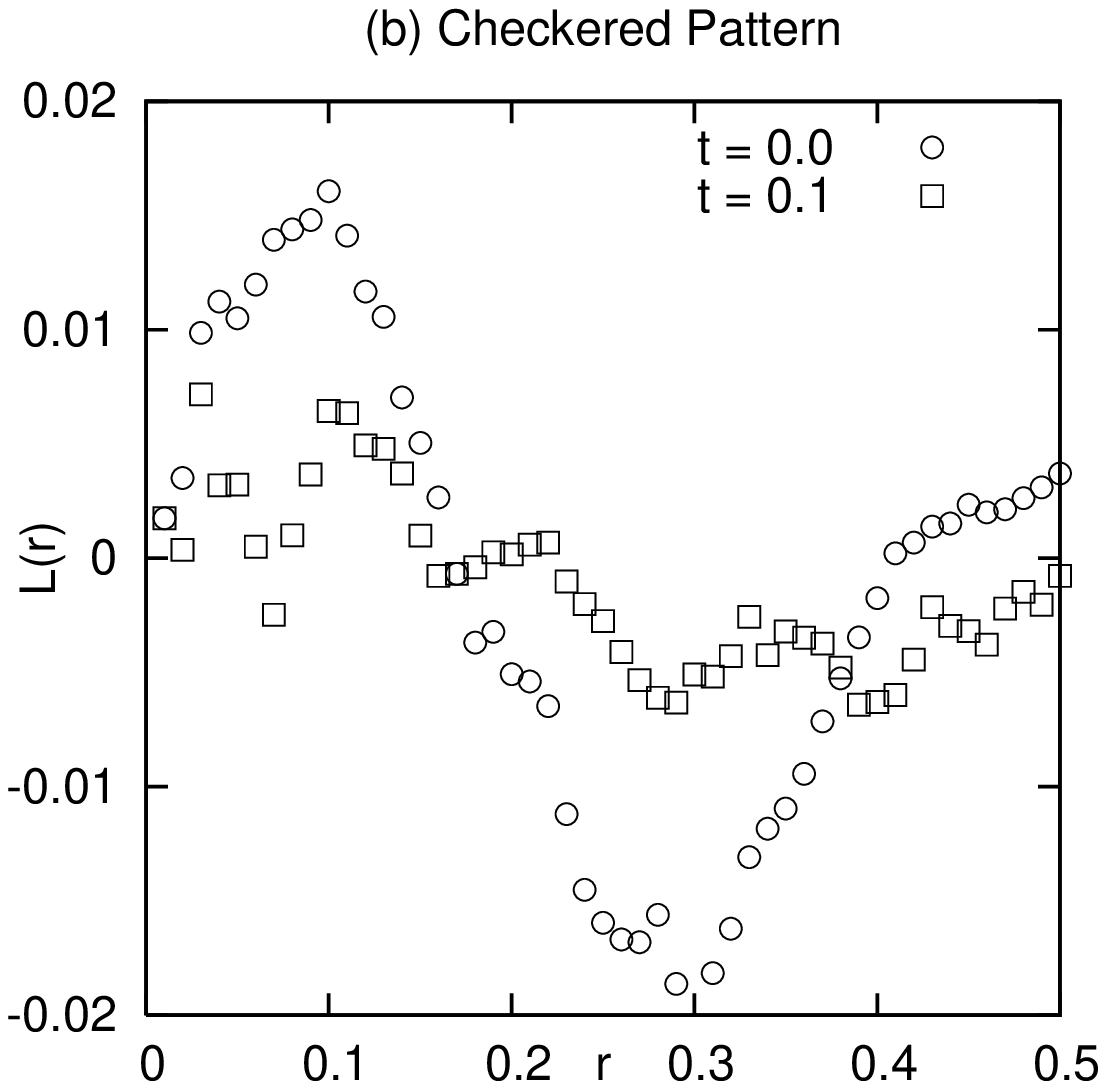}}
\scalebox{0.33}{\includegraphics{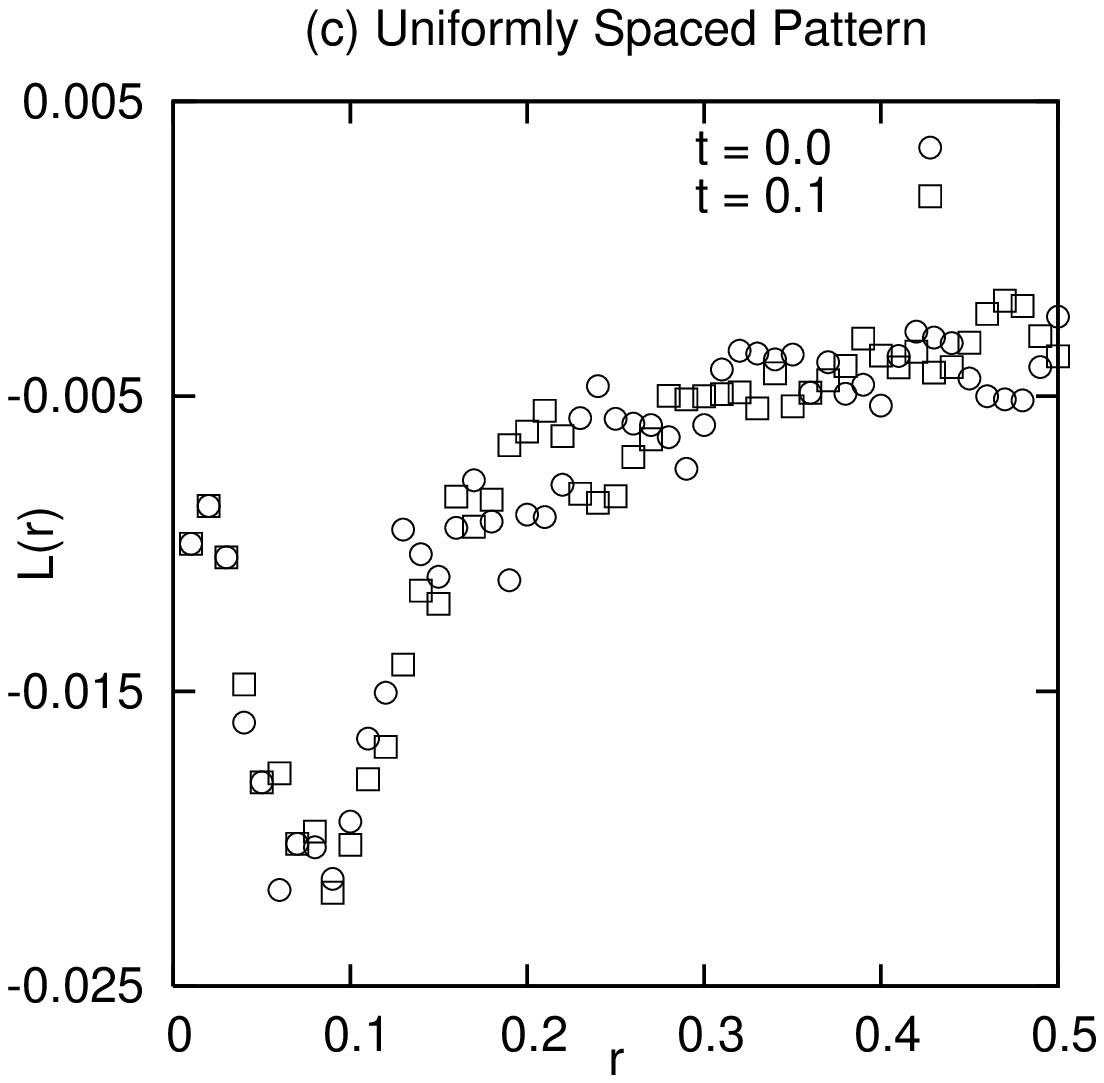}}
\scalebox{0.33}{\includegraphics{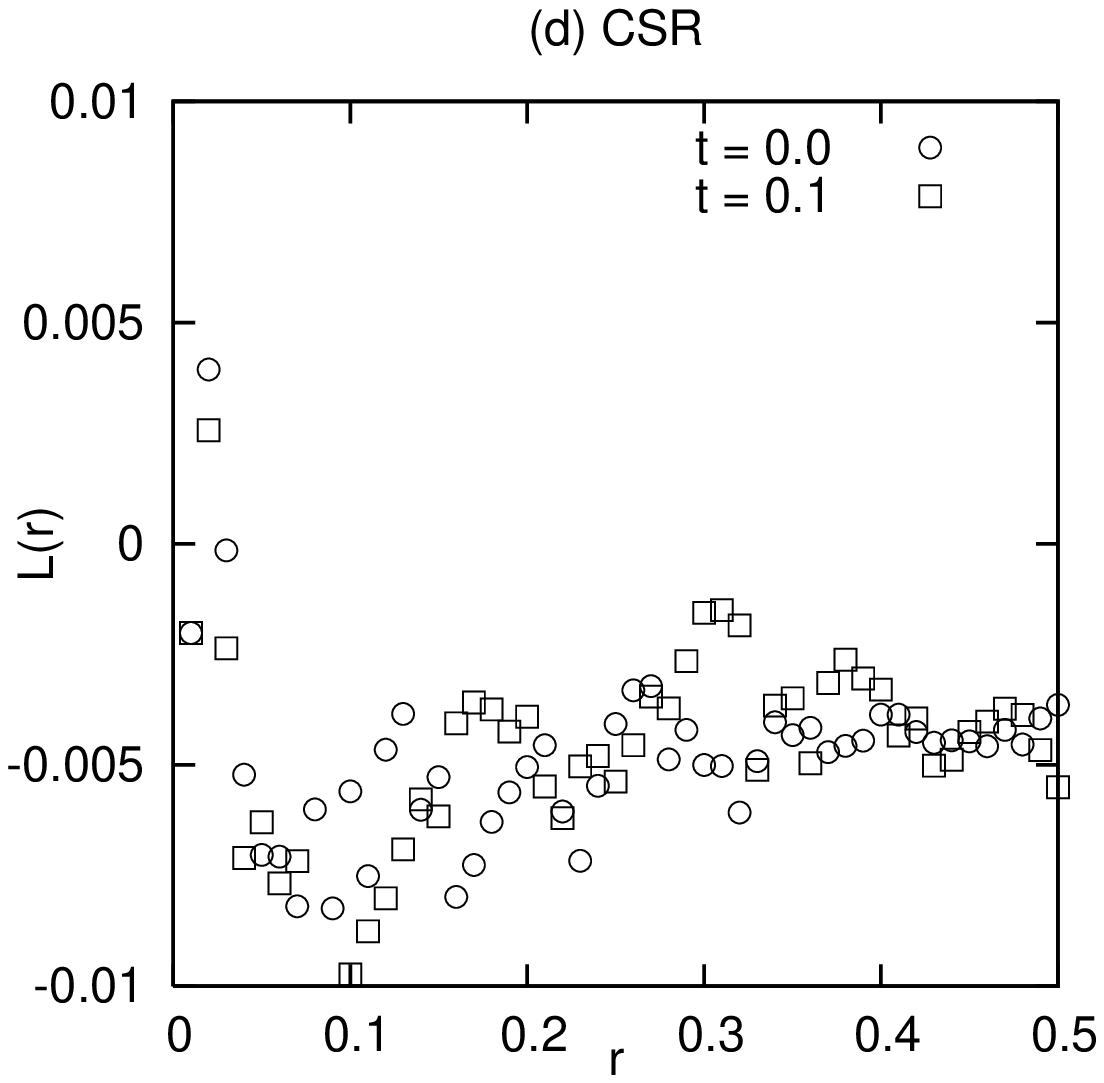}}
\caption{The $L$ functions. (a) corresponds to (I), 
(b) to (II), (c) to (III), and (d) to (IV)}
\end{center}
\end{figure}
%Fig.2

%Fig3
\begin{figure}
\begin{center}
\scalebox{0.33}{\includegraphics{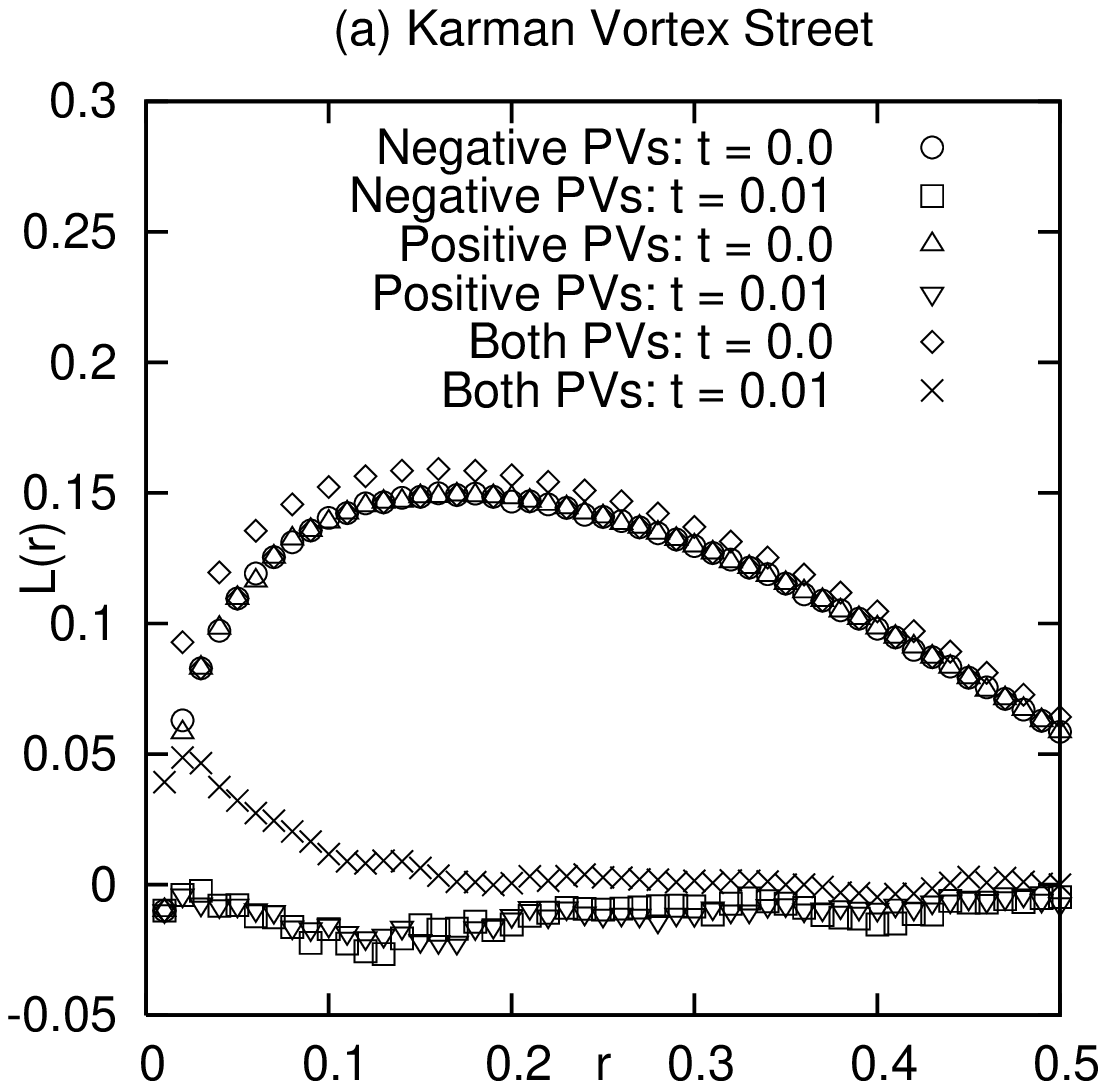}}
\scalebox{0.33}{\includegraphics{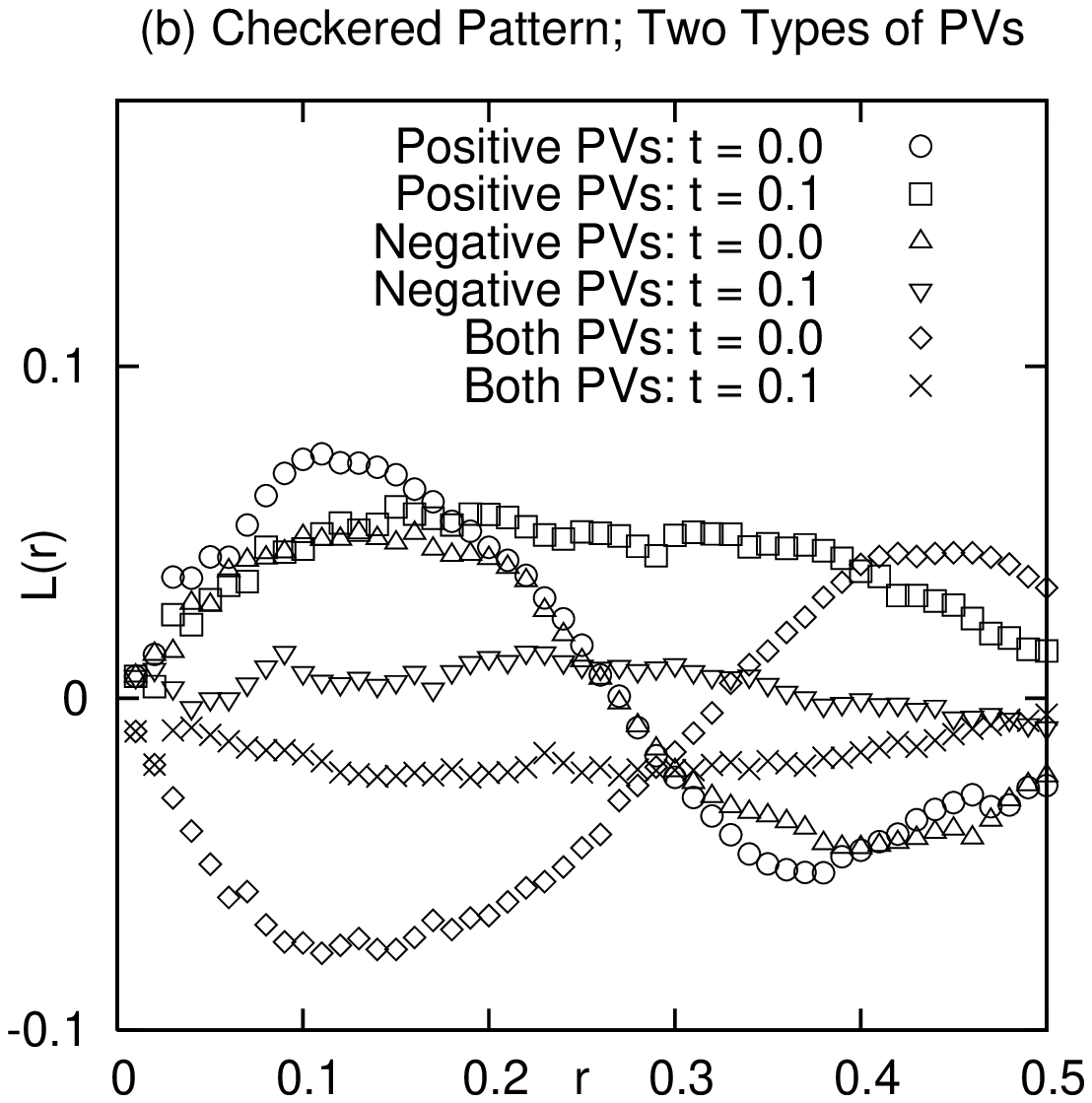}}
\scalebox{0.33}{\includegraphics{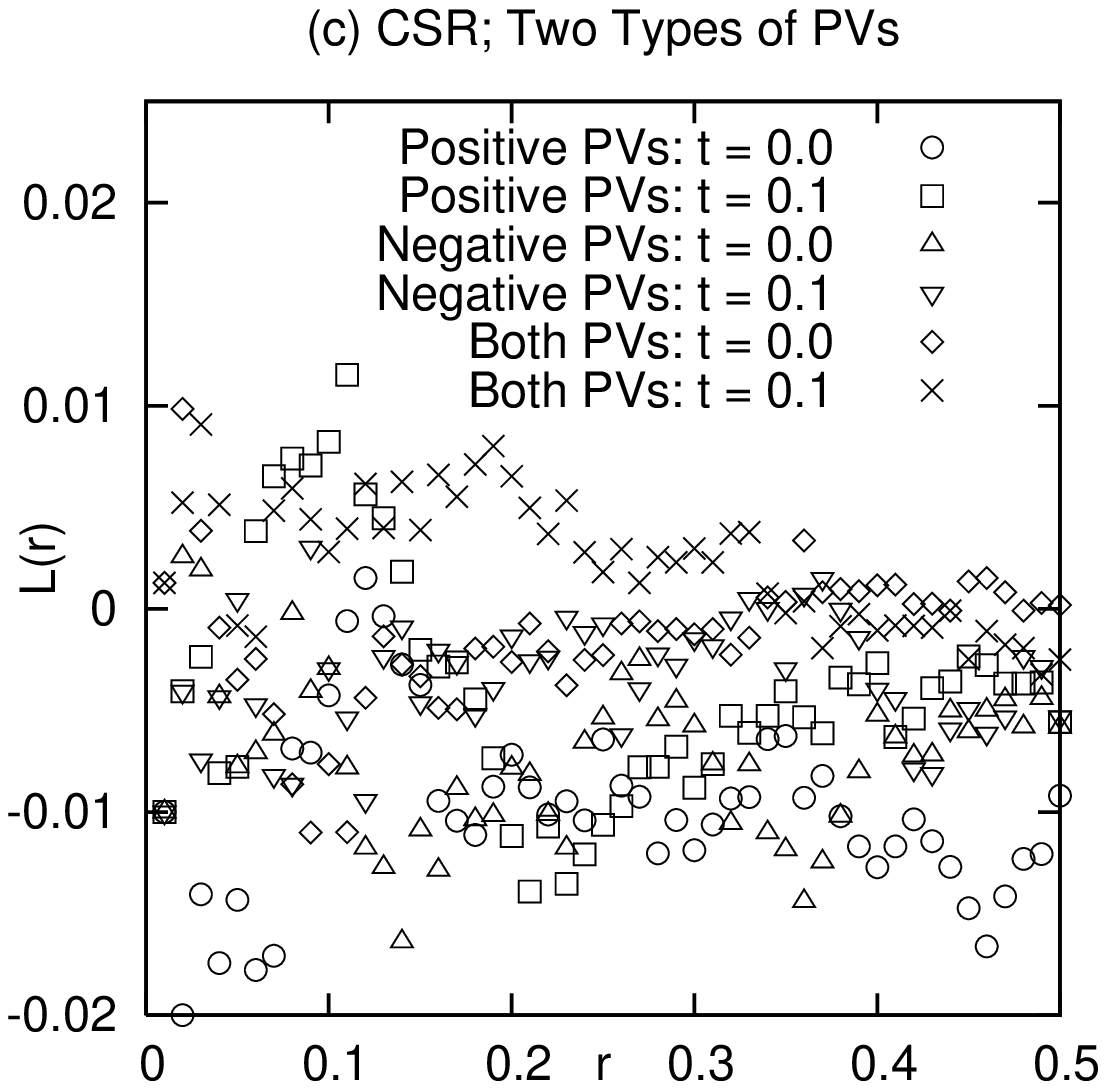}}
\scalebox{0.33}{\includegraphics{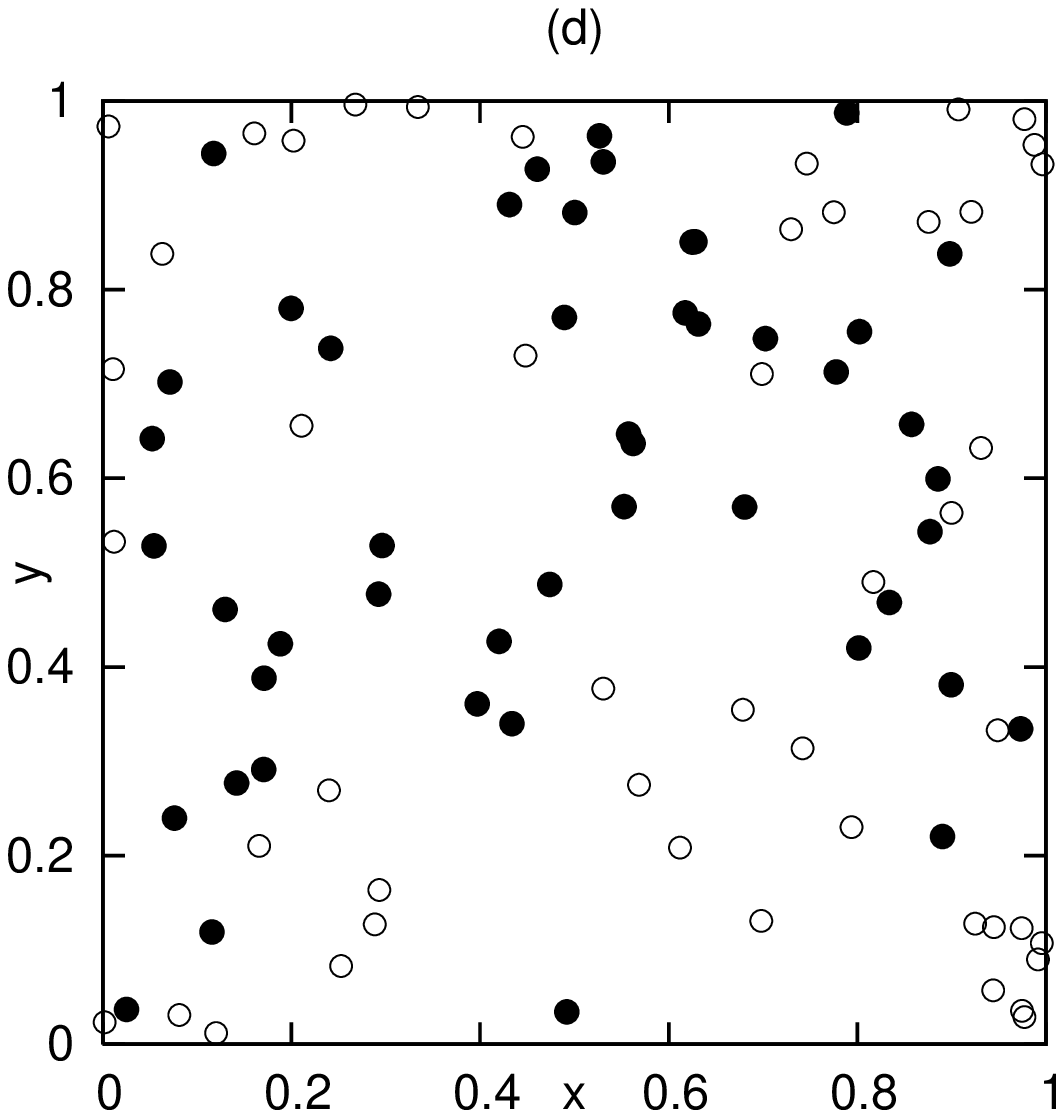}}
\caption{The $L$ functions. (a) corresponds to (V), 
(b) to (VI), and (c) to (VII). (d) shows the distribution 
of PVs for (VI) at $t=0.1$. 
White (black) circles denote positive (negative) PVs.} 
\end{center}
\end{figure}
%Fig3
%\fi

Second, we consider Case (II) where the initial PVs are located 
randomly in eight segments revealing a checkered pattern.  
$L(r)>0$ for $0<r<0.15$ implies that the PVs are clustered, 
while $L(r)<0$ for $0.15<r<0.4$ means that 
they are uniformly spaced at larger scales. 
We observe that this tendency remains at $t=t_f$, although 
it becomes somewhat weak and an additional oscillatory behavior 
is observed. 

Third, Case (III) initially has uniformly spaced PVs. 
$L(r)$ has a negative value and a minimum at $r\approx 0.06$. 
We observe that $L(r)$ is almost the same during $t \in [0,t_f]$.

Fourth, we examine the completely spatially random distribution 
of the PVs at $t=0$ in Case (IV). We observe slightly negative values 
of $L(r)$ at $t=0$ and $t_f$, but we do not see any significant 
differences between the initial and final distributions. 

Next, we consider the system with both positive and negative PVs 
of the same strength and the same numbers $N_1=N_2=N/2$. 
The following three typical cases are examined: 
Case (V) the K\'arm\'an vortex street, 
Case (VI) positive and negative PVs located alternately 
in checkered segments (16 subsquares), and
Case (VII) completely spatially random distribution. 
The initial conditions, the number of PVs, 
$t_f$, and the values of three conserved quantities 
are summarized in Table 2.

First, we consider Case (V) given by the following expression:
%$\mu_j=(-1,1) $, 
%$z_j(0)=(j/N_1 + \epsilon (R+ iR),$ $j/N_2 + 1/N + $ 
%$i h + \epsilon (R+ iR) )$ for $j= (1,\cdots, N_1,$$N_1+1,\cdots, N),$
$(\mu_j, z_j(0))=$ $(-1, j/N_1 \ + \ $ $\epsilon (R+ iR))$
for $j= 1, \cdots, N_1$ and 
$(\mu_j, z_j(0))=$ $(1, j/N_2 + 1/N + $ 
$i h + \epsilon (R+ iR) )$
for $j= N_1+1,\cdots, N$, 
where $\epsilon=10^{-4}$, $R$ denotes random numbers, and the distance 
$h$ between two rows is taken as $1/N$. 
In this case, the negative PVs are first numbered. 
At approximately $t=0.001$, we observe that the two rows begin to break 
with the pairing instability\cite{Saffman}. 

For the two types of PVs, we introduce $K_{lm}(r)$ for 
$(l,m)=(+,+),(-,-)$, and $(+,-)$ as 
\begin{equation}
K_{lm}(r) = (\lambda N_1)^{-1}
\Sigma_{i}\Sigma_{j}
\theta( r- | \vec{x}_i-\vec{x}_j |), 
\label{eq01a}
\end{equation}
where the sum is for $i,j=1, \cdots, N_1$ except for $j= i$ 
if $(l,m)=(+,+)$. 
The case $(l,m)=(-,-)$ is similar. 
The sum for $(l,m)=(+,-)$ is both for $i=1, \cdots, N_1$, 
$j=N_1+1, \cdots, N$ and $i=N_1+1, \cdots, N$, 
$j=1, \cdots, N_1$. 
We define $L_{lm}(r)$ from $K_{lm}(r)$ similar to 
Eq. (\ref{eq02}). 
We observe strong clustering for $(l,m)=(+,-)$
at $t=0.01$ corresponding to the pairing instability. 

Second, for Case (VI), we observe the asymmetry, i.e., 
$L_{++}(r) \gg L_{--}(r)$ for all $r$ at $t=0.1$. 
The initial clustering is incidentally 
stronger for positive PVs than the negative ones. 
This fact can be regarded as clustering to a single vortex 
at the largest scale. 
There is also a significant void where positive (or negative) 
PVs do not exist at $t=t_f$. 
We can also consider the initial {\it fractal} 
distribution like a Sierpinski's gasket, which shows 
clustering at large scales. 

Finally, we consider the CSR distributions of two 
types of PVs as that in Case (VII). 
A remarkable feature in this turbulent situation 
is that there are several pairs of positive and 
negative vortices moving linearly at a 
velocity of $\kappa/4h\pi$, where $2h$ is the distance 
between two PVs. 
Since the pair is surrounded by a number of 
other isolated PVs, however, 
the moving direction is bent by a third vortex 
when they cross each other. 
Moreover, if the collision is nearly head-on, 
a vortex of the pair with a sign opposite to 
the third target vortex replaces its partner with the latter 
and then continues to move linearly again. 
An exact analysis of such scattering of three PVs
in an unbounded domain was already given by 
Aref (1979)\cite{Aref1979}. 

The examples of scattering and recoupling of three PVs 
in a periodic box are given by an initial location 
$(z_1,z_2,z_3)=(L+i L,(L+d+h)i,(L+d-h)i)$ with $L=1/2$. 
A pair of vortices 2 and 3 is initially approaching vortex 1. 
Figure 4a shows their trajectories when $h=0.02$ 
and $d=\pm 0.02, \pm 0.01,0.04,0.08,0.16,$ and $0.32$. 
The final time $t_f$ is 0.04 except for $t_f=0.1$ for $d=0.16$, 
$t_f=0.05$ for $d=0.32$, and $t_f=0.06$ for $d=-0.01$. 
Recoupling is observed for $d=0.01, 0.02, 0.04, 0.08$, 
and $0.16$. 
The dependence of a $\pi$-normalized 
scattering angle $\delta \phi/\pi$ measured by 
the moving direction of $z_3(t=0.04)$ for $h=0.02$ 
in a periodic box is shown in Figure 4b. 
The recoupling of PVs in an unbounded plane 
with $L \rightarrow \infty$ \cite{Aref1979} occurs if 
$0<d/h<9$. 
On the other hand, the present simulation in a periodic box 
with $h=0.02$ shows 
a shift of the range $d$ for recoupling as 
$-0.5\lesssim d/h \lesssim 8.5$, 
although this range may vary as 
$h$ is changed in the case of periodic BCs. 
In a GIF animation, 
successive scattering and recoupling, similar to the chaos in 
a billiard system, are clearly observed.
The existence of such vortex pairs may play a crucial 
role in stirring assemblies of PVs. 

Since the average distance of randomly located PVs 
is $l \sim N^{-1/2}$ and the strength $2\pi$ is fixed, 
the typical velocity and eddy turnover time are 
$v \sim N^{1/2}$ and $t_e \sim 1/N$, respectively. 
Denoting the smallest distance of the vortex pair 
by $\alpha l$, its velocity is $V \sim 1/\alpha l$. 
Because of the recoupling condition, 
the cross section of the scattering is approximately 
$ \sigma_c \sim d \sim 10 \alpha l $ 
and the area swept by the pair during $t_e$ is 
$S \sim d V t_e \sim 10/N$. 
Therefore, the condition for the scattering to occur 
in $t_e$ is $N \sim 10$ since $S \sim 1$, the size of the square. 
If $N=100$, $t_e \sim 0.01$ and 
one pair will be scattered approximately 10 times in a 
numerical simulation in the time interval 
$0 \le t \le 0.1 \sim 10 t_e$.

The $L_{lm}(r)$ function becomes slightly positive for $(l,m)=(+,-)$, 
which also indicates that the pairs of positive and negative 
PVs survive until $t=0.1$. However, the absolute values are much 
smaller than the initially clustered cases. 
To clearly observe the spontaneous clustering, longer simulations 
may be required. 
We also investigated the probability distribution function 
of velocity circulation, which is studied in Umeki 
(1993)\cite{Umeki1993} for a 3D turbulence\cite{Migdal}. 
A similar approach in the point process theory is called 
the Quadrat method\cite{Cressie}. 

In summary, a method to simulate the motions of PVs with 
periodic BCs is described. Several numerical examples are illustrated 
and the clustering of PVs with different conditions 
is examined by the $L$ function. 

The author is grateful to Professor Yamagata for support 
through his research on fluid dynamics over several years.

\begin{table}
\caption{Single type of PVs}
\begin{center}
\begin{tabular}{|c|c|c|c|c|c|}
\hline
Case& $ N $ & $t_f$ & H & $I_x$ & $I_y$\\
\hline
I & 100 & 0.01 & 17948 & 50.5 & 0 \\
\hline
II & 96 & 0.1& 11973 & 47.872 & 47.923 \\
\hline
III & 100 & 0.1& 12796 & 49.968 & 50.020\\
\hline
IV & 100 & 0.1 & 12886 & 49.687&  50.206\\
\hline
\end{tabular}
\end{center}

\end{table}
\begin{table}
\caption{Two types of PVs}
\begin{center}
\begin{tabular}{|c|c|c|c|c|c|c|}
\hline
Case&  $N_{1,2}$ & $t_f$ & H & $I_x$ & $I_y$ \\
\hline
V & 50 & 0.01 & -415.31 & -0.50019 & -0.49976 \\
\hline
VI & 48 & 0.1 & 369.60 & -12.176 & -11.081 \\
\hline
VII & 50 & 0.1 & -178.31 & -1.9433 & 2.3455 \\
\hline
\end{tabular}
\end{center}
\end{table}

\begin{figure}
\begin{center}
\scalebox{0.2}{\includegraphics{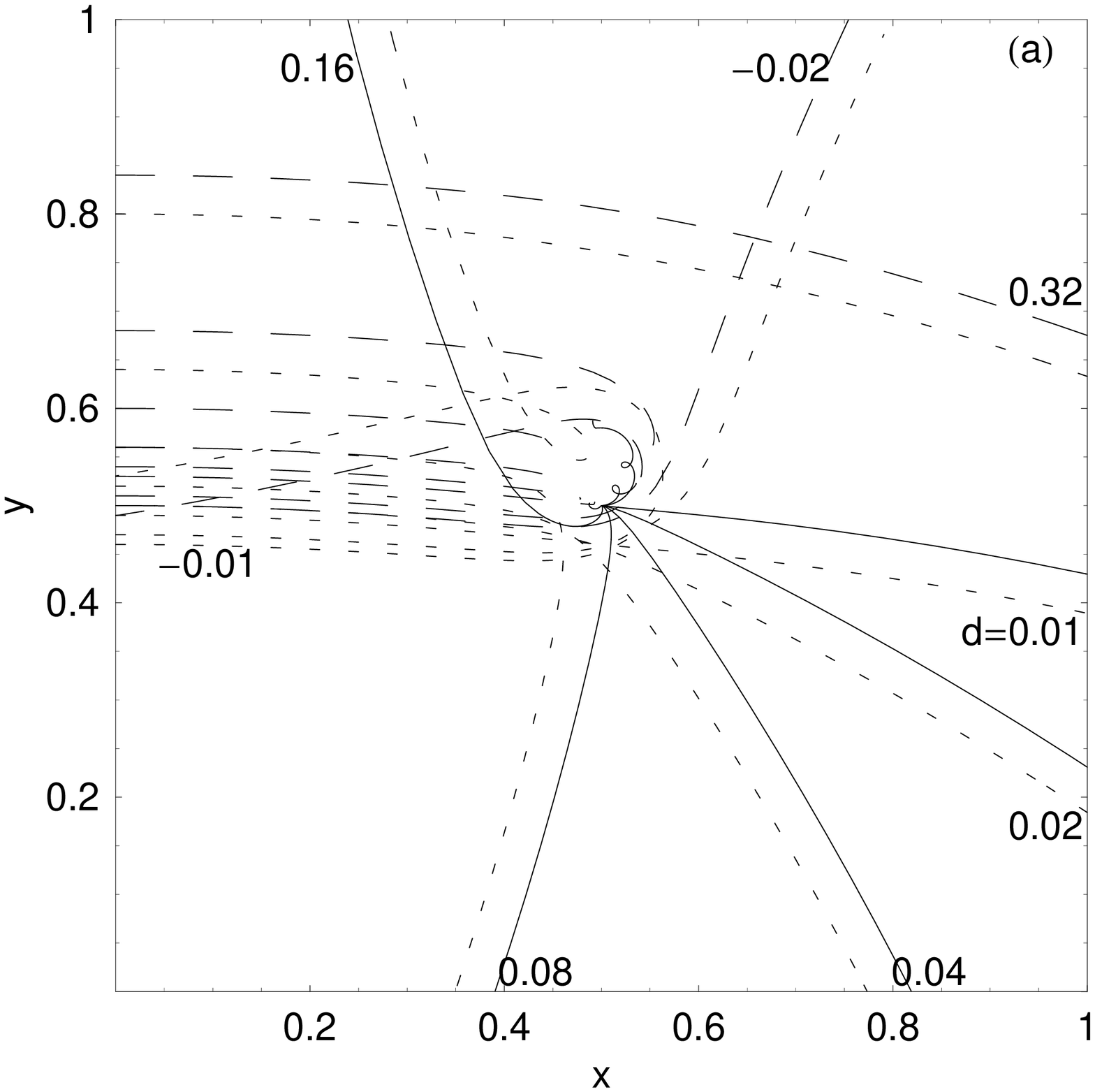}}
\scalebox{0.26}{\includegraphics{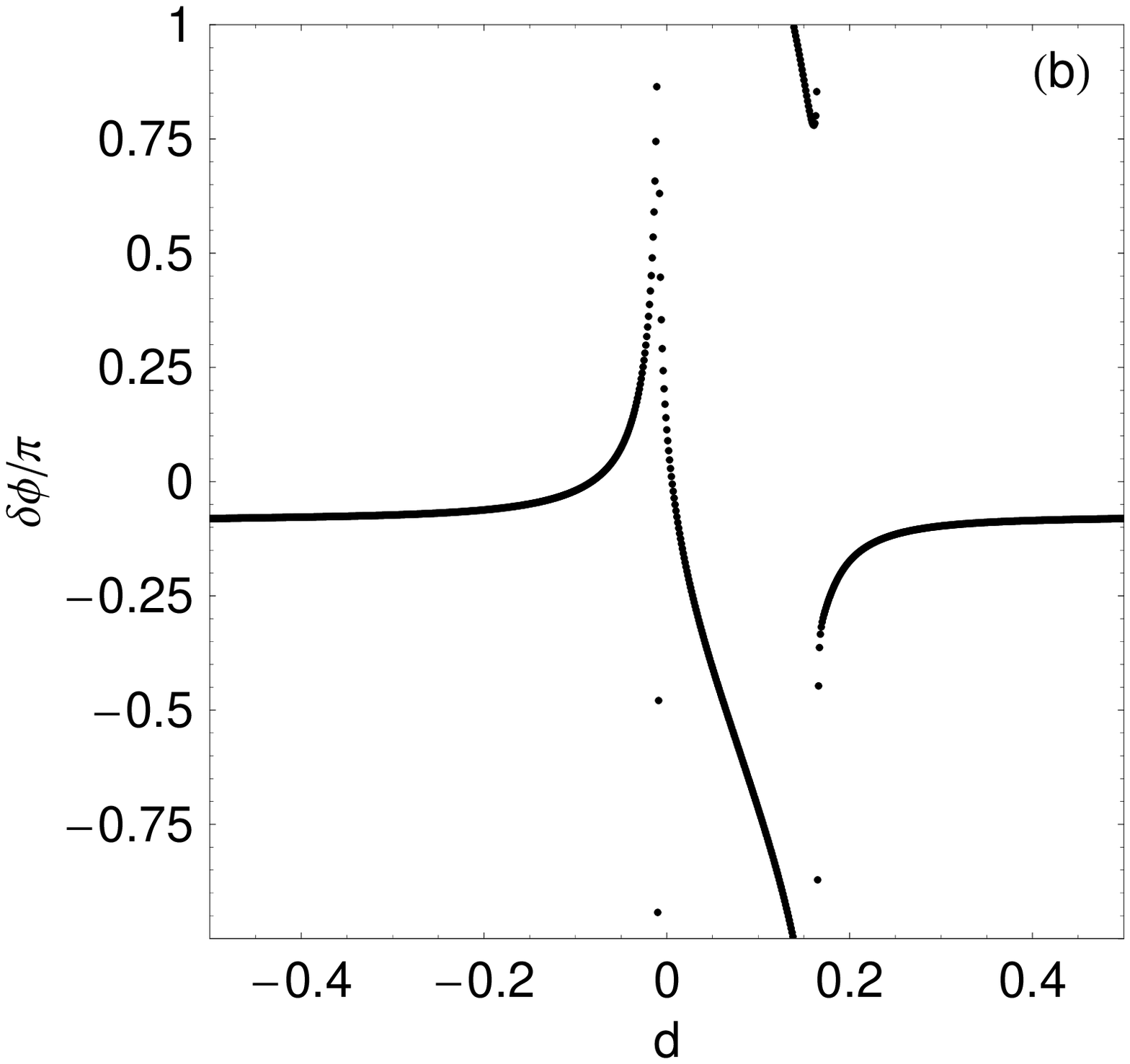}}
\end{center}
\caption{(a) Trajectories of three scattering and recoupling PVs 
for various values of $h$. 
Solid, dashed, and dotted curves denote $z_1$, $z_2$, and $z_3$, 
respectively. 
(b) The $\pi$-normalized scattering angle $\delta \phi/\pi$ versus $d$ 
for $h=0.02$.
}
\end{figure}


\begin{thebibliography}{99} 

\bibitem{tka1} V. K. Tkachenko: {\it Sov. Phys. JETP}
\textbf{22} (1966) 1282.
\bibitem{tka2} V. K. Tkachenko: {\it Sov. Phys. JETP}
\textbf{23} (1966) 1049.
\bibitem{Aref1999} M. A. Stremler and H. Aref: 
{\it J. Fluid Mech.} \textbf{392} (1999) 101.
\bibitem{Saffman} P. G. Saffman: {\it Vortex Dynamics} 
(Cambridge University Press, Cambridge, 1992) Chap. 7. 
\bibitem{Cressie} N. A. C. Cressie: 
{\it Statistics for Spatial Data, Revised Edition} (Wiley, New York, 1993).
\bibitem{Shimatani} K. Shimatani: {\it Jpn. J. Ecology} \textbf{51}
(2001) 87 [in Japanese]. 
\bibitem{Novikov} E. A. Novikov:
{\it Sov. Phys. JETP} \textbf{41} (1976) 937.
\bibitem{Aref1979} H. Aref: {\it Phys. Fluids} \textbf{22} (1979) 393.
\bibitem{Umeki1993} M. Umeki: {\it J. Phys. Soc. Jpn.} \textbf{62} (1993) 3788.
\bibitem{Migdal} A. A. Migdal: {\it Int. J. Mod. Phys.} A \textbf{10} (1994) 1197.

\end{thebibliography}
\end{document}